\documentclass[print,trackchanges]{ati}	
\usepackage{graphicx,times}
\usepackage{lmodern}
\usepackage[sort&compress, numbers,super]{natbib} 
\bibpunct[,]{[}{]}{,}{n}{}{,}			
\usepackage{url}
\usepackage[colorlinks=true,breaklinks=true, linkcolor=red,urlcolor=magenta,citecolor=blue,anchorcolor=green]{hyperref}

\usepackage{booktabs}
\usepackage{amsmath}
\usepackage{amssymb}
\usepackage{caption}
\usepackage{rotating}
\usepackage{CJKutf8}	
\usepackage{fancyhdr}
\usepackage{xcolor}

\renewcommand{\citet}[1]{\citeauthor{#1}(\citeyear{#1})\cite{#1}}	

\let\citep\supercite
\renewcommand{\citet}[1]{\citeauthor{#1}(\citeyear{#1})\supercite{#1}}

\begin{document}

   \title{
  \texttt{Nii-body}: Bayesian Inference of Multiplanet Dynamics via N-body Simulations}


\author{Hong-Fei Jia
   \inst{1}
   \ORCID{0009-0000-3916-3761}
\and Sheng Jin\correspondingAuthor{}
   \inst{1}
   \ORCID{0000-0002-9063-5987}
\and Dong-Hong Wu
   \inst{1}
   \ORCID{0000-0001-9424-3721}
\and Shang-Fei Liu
   \inst{2,3}
   \ORCID{0000-0002-9442-137X}
}
\correspondent{Sheng Jin}	
\correspondentEmail{jins@ahnu.edu.cn}

\institute{Department of Physics, Anhui Normal University, Wuhu, Anhui 241002, China; \email{jins@ahnu.edu.cn}
          \and
          School of Physics and Astronomy, Sun Yat-sen University, Zhuhai 519082, China;
          \and
          CSST Science Center for the Guangdong-Hong Kong-Macau Great Bay Area, Sun Yat-sen University, Zhuhai 519082, China}

   \date{Received:~March 26, 2023;   Accepted:~October 26, 2023;  Published Online:~December 26, 2023; 
   \DOI{ati2024999} }			
   \citeinfo{Jia, H.-F. et al.}\volume{1}\issue{5} \pages{560--566}	
   \StartPage{560} 			
   \MonthIssue{September}		
   \copyrights {2024}     		
   \abstract{
Many exoplanetary systems are multiplanet configurations whose long-term dynamics are governed by N-body gravitational interactions. Consequently, their detection signatures cannot be adequately described by Keplerian orbits.
Accurately interpreting the observational data of these systems---including radial velocity (RV), astrometry, and transit timing variations (TTVs)---requires N-body integration.
Motivated by this need, we developed a Bayesian fitting framework that couples N-body integration with Markov chain Monte Carlo (MCMC) to retrieve the system parameters of multiplanet systems.
The code, named \texttt{Nii-body}, integrates an adaptive Runge--Kutta--Fehlberg 7(8) (RKF78) solver with an automated parallel tempering MCMC algorithm. 
Using simplified synthetic astrometric observations, we evaluated the efficiency and robustness of \texttt{Nii-body}'s N-body orbit retrieval on an idealized two-planet model, demonstrating its potential for future application to real observational data.
The N-body fitting workflow can be readily extended to RV, TTVs, or combined datasets, providing a versatile engine for high-precision orbital inference in multiplanet systems.
}

   \authorrunning{ASTRONOMICAL TECHNIQUES \& INSTRUMENTS }   
   \titlerunning{Jia, H.-F. et al.: ~Prepare a LaTeX Manuscript for ATI }  
   \maketitle
   \setcounter{page}{\Page}	
%
%
\section{Introduction}

Over the past three decades, exoplanet surveys have revealed that planetary systems are a common result of star formation~\citep{Winn2015,Zhu2021,Petigura2013,Mulders2018}. These systems exhibit a wide range of architectures, including multiplicity, orbital spacing, and dynamical structure~\citep{Fang2012,Zhu2021,Lissauer2011,Muresan2024,Mishra2023}. Statistics from the transit detections by the Kepler space telescope have revealed the prevalence of compact, multiplanet systems and have also enabled population-level constraints on occurrence rates and system architectures~\citep{Borucki2010,Batalha2013,Fabrycky2014}.
Ongoing catalog curation and follow-up observations have shifted the exoplanet research from discovery-driven  to characterization-driven~\citep{Winn2015}.
This expanding field of study calls for quantitative, dynamic models that link observed architectures to formation channels, migration histories, and long-term stability.
Such models are particularly important for multiplanet systems, where planet--planet interactions can reshape configurations on secular timescales~\citep{Pu2015,Smith2009}.

Characterizing an exoplanetary system encompasses fitting observational data obtained by several detection techniques from different missions. 
RV monitoring measures the host star's line-of-sight reflex motion and remains a workhorse for constraining planetary masses~\citep{Mayor1995,Lovis2010}.
TTVs encode planet--planet interactions and can constrain planetary masses and eccentricities, particularly in compact and resonant systems~\citep{Holman2005,Lithwick2012}. 
High-precision astrometry measures the sky-plane reflex motion and can break the $m\sin i$ degeneracy by constraining inclination and the full three-dimensional architecture~\citep{Sozzetti2005,Perryman2014,Lindegren2018,Ji2022,Huang2025}.

Bayesian inference provides robust posterior constraints on orbital parameters for time-series data from various detection methods~\citep{Ford2005}.
For multiplanet systems, Bayesian orbital fitting requires generating a time series at every likelihood evaluation, an operation that becomes computationally expensive when N-body integration is used.
Consequently, many widely used tools---e.g. \texttt{EXOFIT}~\citep{Balan2009} and \texttt{ORVARA}~\citep{Brandt2021}---adopt a Keplerian superposition approach in which the orbital motion of a multiplanet system is modeled as a linear sum of independent two-body orbits. 
This approximation is adequate for short data spans or when planet-planet interactions are weak. 
However, it fails catastrophically when perturbations accumulate~\citep{Laughlin2001}. 
This situation is encountered in resonant, or near-resonant, configurations and in tightly packed architectures~\citep{Agol2005}. 
In these regimes, mutual planetary perturbations can drive long-term drifts in the observables from RV, astrometry, and TTVs, which invalidate the Keplerian superposition method. 
In such cases, accurate characterization of the system parameters can only be achieved using a self-consistent N-body forward model.

To obtain rigorous Bayesian fits of multiplanet systems, several codes now incorporate direct N-body integration. 
Examples include \texttt{PlanetPack}~\citep{Baluev2013,Baluev2018}, \texttt{Exo-Striker}~\citep{Trifonov2019}, and \texttt{orbitize!}~\citep{Blunt2020}, the latter providing seamless interfaces to \texttt{REBOUND} and its high-precision IAS15 integrator~\citep{Rein2012,Rein2015}.
In this work, we introduce \texttt{Nii-body}\footnote{https://github.com/shengjin/nii-body.git}, an open source orbit fitting code that integrates the RKF78 integrator with an automatic parallel tempering Markov chain Monte Carlo (MCMC) algorithm~\citep{Jin2022,Jin2024}.
Based on retrieval tests using idealized synthetic astrometric observations, \texttt{Nii-body} is able to recover orbital parameters from the full parameter space through blind searches, demonstrating its potential for future application to real observational data.

This paper is organized as follows. 
Section \ref{sec:model} describes the implementation details of our N-body MCMC fitting algorithm.
Section \ref{sec:nbody} benchmarks the accuracy and speed of the internal RKF78 integrator of  \texttt{Nii-body}  with the \texttt{REBOUND} package.
Section \ref{sec:kep9peak} validates the retrieval performance of \texttt{Nii-body} by fitting synthetic astrometric observations of the Kepler-9 system.
Section \ref{sec:sum} provides a brief summary.


\section{Bayesian N-body fitting}
\label{sec:model}

\subsection{RKF78 Orbit Integrator}

Although Keplerian superposition provides an efficient method for calculating the orbits of multiplanet systems due to its computational efficiency, this approximation can cause significant long-term drifts in the phase and amplitude of observables, particularly in systems with strong planet-planet interactions and resonant or near-resonant architectures~\citep{Laughlin2001,Holman2005,Lithwick2012}.

\begin{figure*}[!htb]
    \centering
    \includegraphics[width=0.9\textwidth]{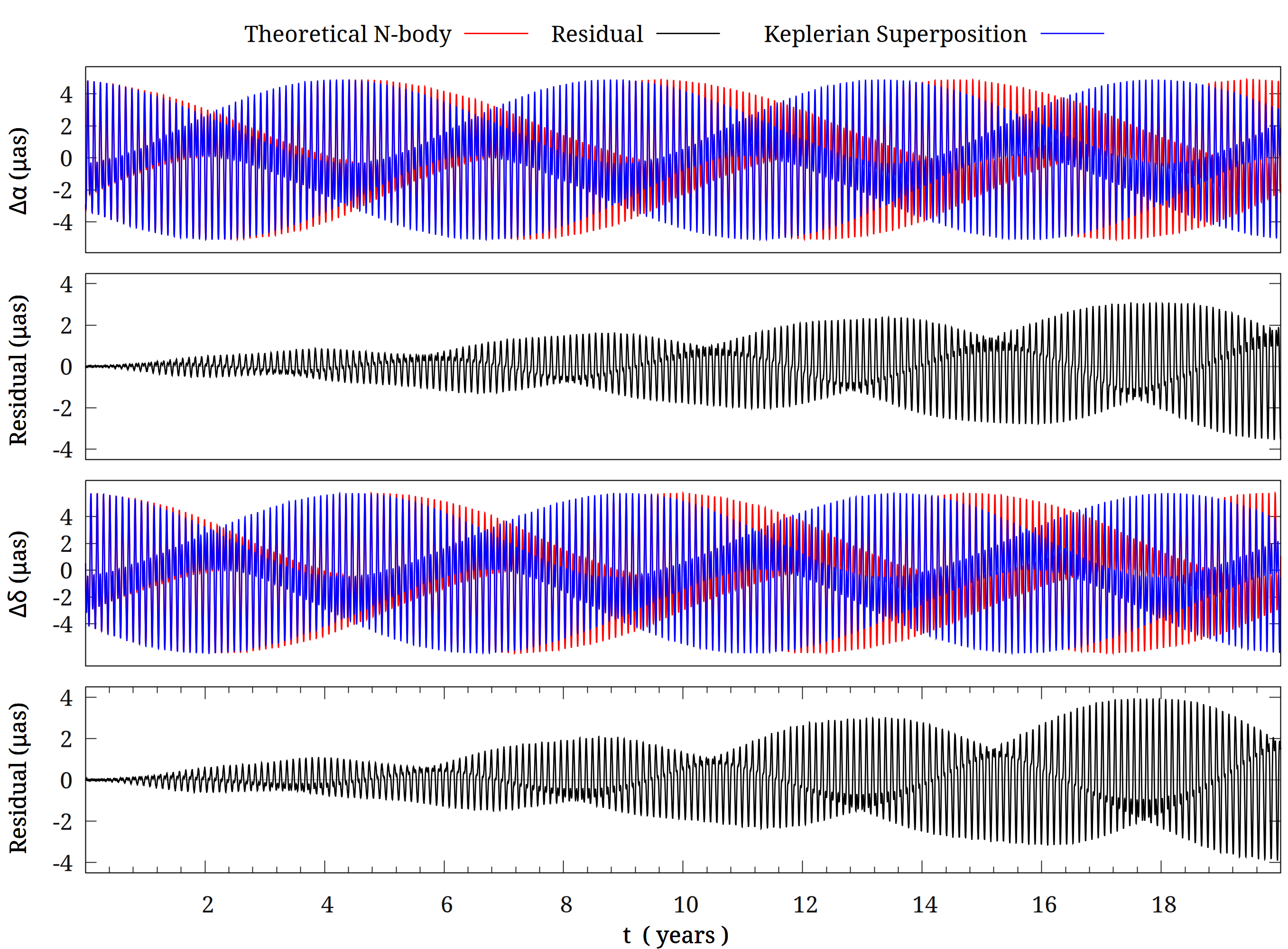}
    \caption[]{Synthetic astrometric signals of the host star in the Kepler-9 system, influenced by gravitational perturbations from two planets in a 2:1 resonance. 
    The first and third panels show the time series of the astrometric offsets of right ascension ($\Delta\alpha$) and declination ($\Delta\delta$), respectively.
    The red curve is calculated using N-body integrations that account for the mutual interactions between the planets, 
    while the blue curve is obtained by superposing two independent Keplerian signals.
    The second and fourth panels show the residuals between the two models (defined as N-body minus Keplerian) for $\Delta\alpha$ and $\Delta\delta$, respectively, thereby making the dynamical perturbation signal more clearly visible.
    The system parameters used to generate the synthetic signals are listed in Table~\ref{tab1}.}
    \label{fig1}
\end{figure*}

Figure~\ref{fig1} compares the synthetic astrometric observations of the host star's wobble in the two-planet Kepler-9 system, as calculated using N-body integrations that account for the mutual interactions between the planets, with those obtained by superposing two independent Keplerian signals (i.e., assuming each planet orbits the star independently, without interactions with the other).
The parameters used to generate the synthetic astrometric observations of the Kepler-9 system are listed in Table~\ref{tab1}.
The Kepler-9 system features two planets, Kepler-9 b and Kepler-9 c, in a 2:1 mean motion resonance, resulting in strong gravitational perturbations between them.

As shown in Figure~\ref{fig1}, the signal obtained via Keplerian superposition is inadequate for reproducing the long-term drift of the system, illustrating the importance of N-body integrations for accurately interpreting the observables of multiplanet systems. 
Note that the magnitude of the astrometric signals is far below the precision of the current GAIA mission\citep{Lammers2026}. The main purpose of this test case is to assess the retrieval capability of N-body dynamics of the \texttt{Nii-body} code. Such capability is essential for future high-precision missions, including concepts such as THEIA~\citep{Malbet2016,Theia2017} and CHES~\citep{JI2020}, for which dynamical signatures at the microarcsecond level are expected to be achievable.

\begin{table}
\caption[Kepler-9 synthetic parameters]
{Parameters used to generate the synthetic astrometric signals of the Kepler-9 system, taken from the TTVs fitting results~\citep{Freudenthal2018}.}
\footnotesize
\setlength{\tabcolsep}{9pt}
\renewcommand{\arraystretch}{1.2}
\begin{tabular}{ccccc}
\hline
\hline
Planet & $M_{\rm p}$ ($M_\oplus$) & $P$ (days) & $e$ & $i$ (deg) \\
\hline
Kepler-9 b & 41.71 & 19.247 & 0.0638 & 88.94 \\
Kepler-9 c & 30.79 & 38.944 & 0.0680 & 89.18 \\
\hline
\end{tabular}
\label{tab1}
\end{table}
\normalsize

\texttt{Nii-body}  integrates the gravitational N-body dynamics with an embedded RKF78 solver that has local-error tolerance and adaptive step-size control~\citep{Fehlberg1968,Dormand1978}. 
For each object, we track its three-dimensional state vector described by
\begin{equation}
\boldsymbol{\xi}_i = (x_i, y_i, z_i, v_{x,i}, v_{y,i}, v_{z,i})
\end{equation}
where $x_i, y_i, z_i$ give the  barycentric position and $v_{x,i}, v_{y,i}, v_{z,i}$ give the velocity of the $i$-th object.

The RKF78 solver then integrates the standard Newtonian $N$-body equations of motion in barycentric coordinates,
\begin{equation}
\frac{d\boldsymbol{r}_i}{dt}=\boldsymbol{v}_i,
\qquad
\frac{d\boldsymbol{v}_i}{dt}
=
\sum_{j\ne i} m_j
\frac{\boldsymbol{r}_j-\boldsymbol{r}_i}{|\boldsymbol{r}_j-\boldsymbol{r}_i|^3},
\end{equation}
where $\boldsymbol{r}_i=(x_i,y_i,z_i)$ and $\boldsymbol{v}_i=(v_{x,i},v_{y,i},v_{z,i})$ are the barycentric position and velocity vectors of the $i$th body, respectively. In the code implementation, the Cartesian states are shifted to the system barycenter before integration, and the equations are solved in code units with $G=1$.

Our code provides conversion libraries between Cartesian coordinates and the Keplerian elements, specifically $(P, e, i, \omega, \Omega, M_0)$, where $P$ is the orbital period, $e$ the eccentricity, $i$ the inclination, $\omega$ the argument of periastron, $\Omega$ the longitude of the ascending node, and $M_0$ the mean anomaly. 
Both representations can be output on demand during the integration, enabling the N-body integration results to be translated into various synthetic observables without external preprocessing, such as RV, astrometry, TTVs.

The RKF78 orbit integrator is implemented in the C programming language to facilitate easy combination with the automatic parallel tempering MCMC algorithm~\citep{Jin2024}, which will be detailed in the next section.
During the integration, lengths are expressed in astronomical units (AU), masses in solar masses ($M_{\odot}$), and time is converted to a dimensionless variable of $t = 2\pi \cdot {\rm year}$, resulting in a gravitational constant of $G=1$ in the code units.

\subsection{Synthetic Observations}

To convert three-dimensional orbital motion into observable quantities of a planetary system, \texttt{Nii-body} implements internal modules that generate synthetic  astrometric and RV signals directly from $N$-body integrations. 
Our synthetic model neglects the difference between the barycenter and the photocenter, as well as the contribution of  planetary companions to the measured photocenter. The astrometric signal is therefore given entirely by the reflex motion of the host star.

\begin{figure}[!htb]
    \centering
    \includegraphics[width=\columnwidth]{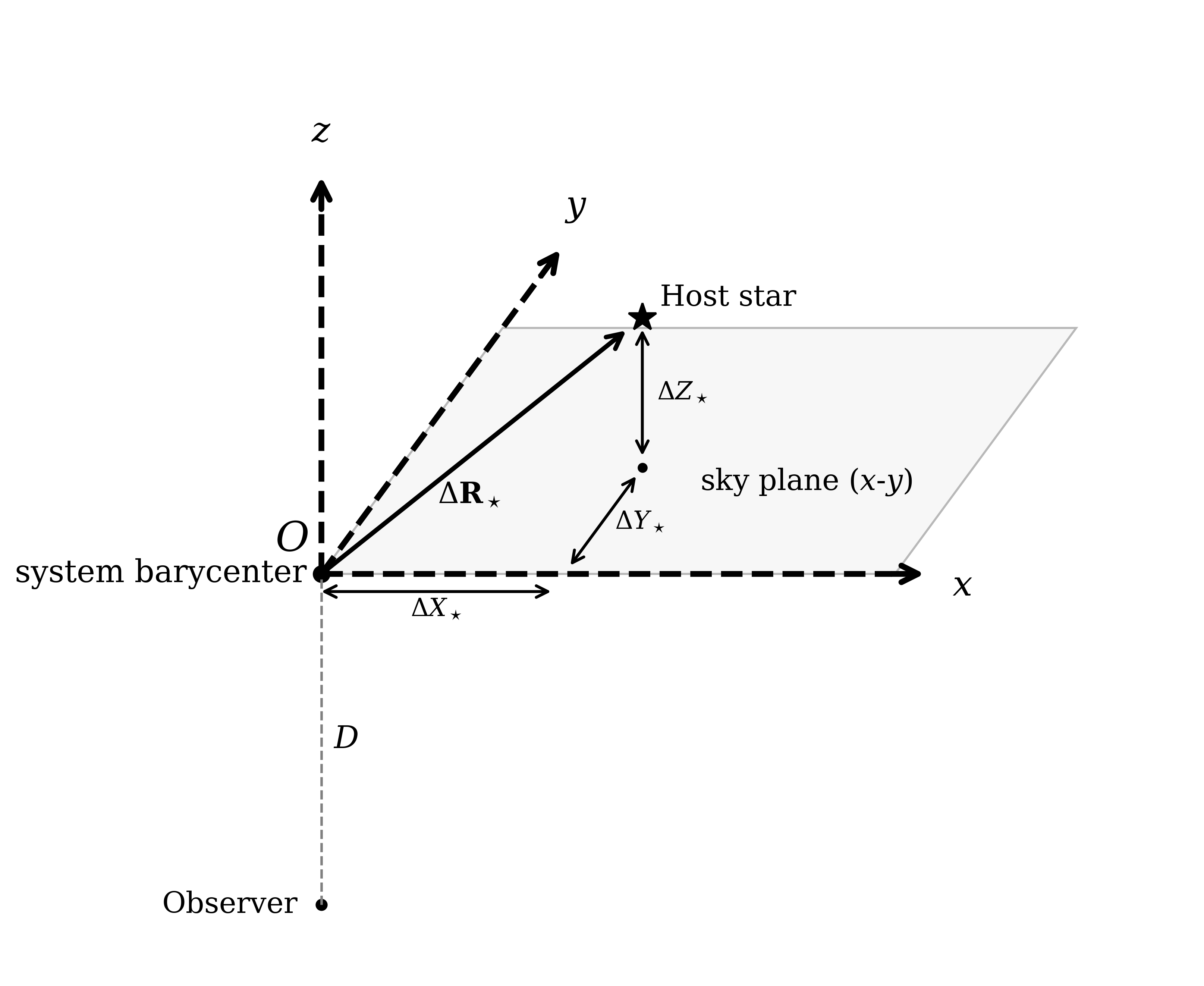}
    \caption{Schematic illustration of the coordinate system used for the synthetic astrometric and radial velocity observables. The origin $O$ is placed at the system barycenter. The host has a barycentric displacement described by $\Delta X_\star$, $\Delta Y_\star$ , and $\Delta Z_\star$. The $x$--$y$ plane represents the plane of the sky, and the positive $Z$ axis points from the observer toward the system barycenter, so that the observer is located at $z=-D$.}
    \label{fig2}
\end{figure}

Figure~\ref{fig2} illustrates the coordinate system and the geometry underlying the synthetic observations.
For synthetic astrometric observations, the three-dimensional barycentric position of the host star, as determined by N-body integration, is projected onto the right ascension and declination plane of the sky, simulating the exact viewing geometry of a real astrometric measurement.
Let the stellar displacement from the system barycenter be the vector $(\Delta X_\star, \Delta Y_\star, \Delta Z_\star)$, and place the observer along the negative $z$ axis at a distance $D$ from the system barycenter. Under the small-angle approximation, the astrometric offsets are given by
\begin{equation}
\Delta \alpha(t) \simeq \frac{\Delta X_\star(t)}{D}, \qquad
\Delta \delta(t) \simeq \frac{\Delta Y_\star(t)}{D},
\end{equation}
where $D$ is the distance from an observer on Earth, 
$\Delta \alpha(t)$ represents the astrometric offset in right ascension, $\Delta \delta(t)$ represents the offset in declination.
The unit of astrometric wobbles in the code is  microarcseconds ($\mu$as).
Our simple astrometric model neglects parallax, proper motion and any bulk drift of the system. It considers only the barycentric reflex motion of the host star. 
Note that, due to this omission, our work is intended only as a proof-of-concept for N-body orbit retrieval of theoretical two-planet perturbations, under strong simplifications that omit important parameters such as parallax, proper motion, and others. In realistic applications, we should perform simultaneous fitting of these parameters together with the planetary signal to resolve additional degeneracies.

For synthetic RV observations, \texttt{Nii-body} uses the time derivative of the stellar barycentric displacement along the line of sight:
\begin{equation}
v_{\rm RV}(t) = -\,\frac{d\,\Delta Z_\star}{dt}.
\end{equation}
We neglect the contribution from the Earth's motion in the calculation of the RV signals.

\texttt{Nii-body} therefore enables a self-consistent numerical description of N-body gravitational interactions, facilitating the accurate modeling of the astrometric and RV observations of multiplanet systems.

\subsection{MCMC Fitting}

N-body integrations produce synthetic astrometric and RV signals that are compared with observed data to compute the likelihood of a given parameter set.
Likelihoods computed for the multitude of parameter sets explored by MCMC are combined with the priors to map the full posterior distribution, which is the cornerstone of Bayesian parameter inference.
Finally, the marginal means (or modes) of the posterior distribution provide the maximum-a-posteriori (MAP) estimates of the orbital parameters.
We adopt uniform priors for all model parameters.
Assuming we have obtained $N$ measurements of a star's astrometric motion over time, the likelihood function for the observed $2N$ values of $\Delta \alpha(t)$ and $\Delta \delta(t)$, given a parameter set $\theta$, is computed as follows:

\begin{equation}
\begin{split}
        \mathcal{L} = \, A\, &\exp \Biggl\{-\sum_{i=1}^{N}\frac{\big[ \Delta \alpha'(t_i) -\Delta \alpha(t_i) \big]^2}{ 2\epsilon_{x}^2 }\Biggr\}\, \\
        & \times \exp \Biggl\{-\sum_{i=1}^{N}\frac{\big[ \Delta \delta'(t_i) -\Delta \delta(t_i) \big]^2}{ 2\epsilon_{x}^2 }\Biggr\}\,
\label{eqn:lik}
\end{split}
\end{equation}
where
\begin{equation}
        A=(2\pi)^{-N}\prod_{i=1}^{N}(\epsilon_{x}^2)^{-1/2}\,\prod_{i=1}^{N}(\epsilon_{x}^2)^{-1/2}\,
\label{eqn:likA}
\end{equation}

where $\Delta\alpha(t_i)$ and $\Delta\delta(t_i)$ denote the synthetic astrometric offsets for the $i$th observation, while $\Delta\alpha'(t_i)$ and $\Delta\delta'(t_i)$ represent the offsets computed from the model corresponding to parameter set $\theta$ at that epoch, $\epsilon_{x}$ represents the adopted $1\sigma$ measurement uncertainty. 
Note that in this synthetic observation model, we assume independent and isotropic Gaussian observational errors in right ascension and declination, such that the same uncertainty $\epsilon_x$ is adopted for both coordinates at all epochs. In real cases, this model should be replaced by a covariance-based formulation to make the methodology more general. We plan to implement such a realistic model in the future when real observational data become available.

\texttt{Nii-body} implements the automatic parallel-tempering MCMC framework given by the \texttt{Nii-C} code~\citep{Jin2024}, which is a high-performance MCMC library written in modern C.
The \texttt{Nii-C} MCMC sampler resolves challenges in sampling high-dimensional or multimodal distributions using two complementary techniques.
First, parallel tempering helps chains escape local modes and enhances exploration of the global parameter space.
Second, an automated control system activates during the initial tuning phase, optimizing each chain’s proposal distributions to ensure a high late-stage  sampling acceptance rate.

In general, the MCMC process is divided into an initial tuning stage and a subsequent production stage. During the tuning stage, the Gaussian proposal widths for each parameter are adjusted automatically to maintain a favorable acceptance rate while the parallel samplers explore the parameter space to locate the global mode.
Once the tuning stage is complete, the sampler reverts to a standard parallel-tempering MCMC with fixed Gaussian proposals to preserve the Markovian property.
Only this later production stage is used for posterior inference.
A detailed illustration of this workflow can be found in \citet{Jin2024}.
We assess whether the MCMC chains have reached convergence using the Gelman-Rubin criterion \citep{Gelman1992}, applied to eight independent runs with different initial random seeds.

In addition to the rapid convergence delivered from its built-in automatic parallel-tempering MCMC engine, \texttt{Nii-body} gains further speed from its complete implementation in the C programming language, ensuring a high computing efficiency.
At least for the idealized synthetic astrometric observations used in this work, \texttt{Nii-body} can recover orbital parameters from the full parameter space through blind searches in most runs, with a computational cost that remains practical despite the intrinsic expense of $N$-body MCMC fitting.

Section \ref{sec:kep9peak} provides examples of the astrometric fitting of a two-planet system with a total of 15 parameters. Using one million MCMC  steps on an Intel i7-12650H CPU laptop, \texttt{Nii-body} achieves convergence in approximately 24 hours.
The computational cost of the present implementation is expected to increase with the number of planets, the number of observational epochs, and the modeled time span, because each likelihood evaluation requires an $N$-body integration over the full fitting baseline together with the computation of synthetic observables at the prescribed epochs. A systematic characterization of this scaling is beyond the scope of the present validation paper and will be investigated in future work. Our current work is intended only to demonstrate the accuracy and practical feasibility of N-body orbit retrieval using \texttt{Nii-body} with a simple two-planet model.

\section{Numeric Tests}

\subsection{N-body Benchmarks}
\label{sec:nbody}

To verify both the accuracy and speed of our RKF78 N-body integrator, we ran a suite of benchmarks against the \texttt{REBOUND} code~\citep{Rein2012}.
The test case is a miniature solar system containing only the Sun, Earth and Jupiter (SEJ). 
We integrated this SEJ configuration for 10,000 years, with a 100-year output cadence, using identical starting conditions and compared the trajectories and execution times produced by \texttt{Nii-body}'s RKF78 integrator and \texttt{REBOUND}'s IAS15 integrator for a set of prescribed error tolerances ($err_{\rm toler}$) applied at each timestep: ${10^{-14},\,10^{-12},\,10^{-10},\,10^{-8}}$.
All benchmarks were executed on a laptop with an Intel i7-12650H CPU.

\begin{figure*}[!htb]
    \centering
    \includegraphics[width=0.9\textwidth]{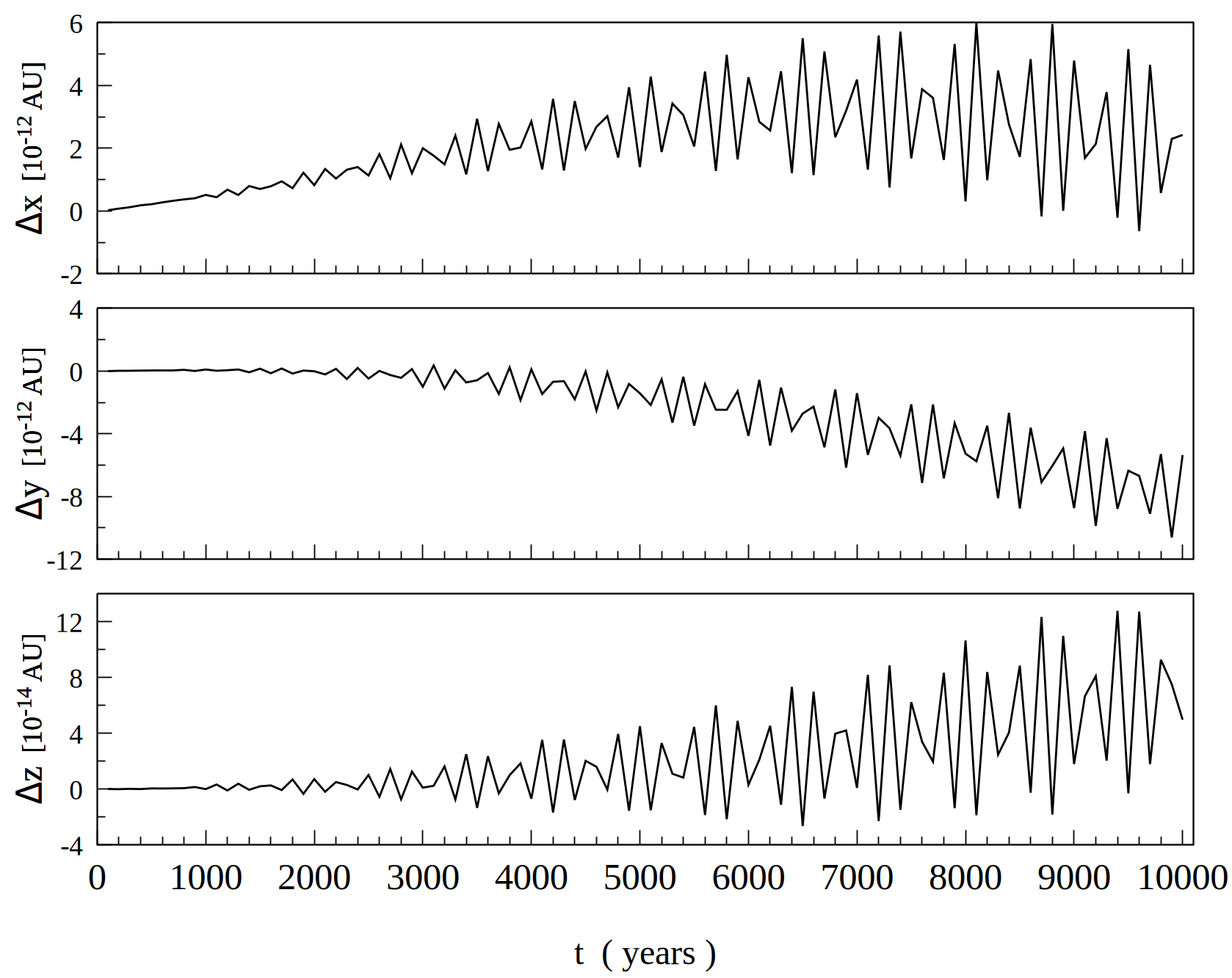}
    \caption{The differences in the $x$, $y$, and $z$ coordinates ($\Delta x$, $\Delta y$ and $\Delta z$) between the trajectories computed by \texttt{Nii-body}/RKF78 and \texttt{REBOUND}/IAS15 for the SEJ benchmark. The residuals remain small over the full 10,000-year integration, confirming the close numerical agreement between the two codes.}
    \label{fig3}
\end{figure*}

Figure~\ref{fig3} compares the $x$, $y$, and $z$ coordinate discrepancies between the trajectories computed by \texttt{Nii-body}/RKF78 and \texttt{REBOUND}/IAS15 for the SEJ benchmark. 
Both codes employ an error tolerance of $10^{-14}$.
The residuals between the two trajectories remain negligibly small throughout the 10,000-year integration, confirming close numerical agreement between the implementations.
To quantify the relative precision of \texttt{REBOUND}'s IAS15 integrator, we computed the Root Mean Square Error (RMSE) as a metric, defined as: 
\begin{equation}
\mathrm{RMSE}
=
\sqrt{
\frac{1}{N}
\sum_{k=1}^{N}
\big\|
\mathbf r_A(t_k)-\mathbf r_B(t_k)
\big\|^2
}.
\label{eq:rmse}
\end{equation}
where $N$ is the number of equally spaced timeseries points $t_k$ ($k=1,\dots,N$) output by the two integrators over 10,000 years, and $\mathbf r_A(t_k)$ and $\mathbf r_B(t_k)$ denote the three-dimensional position vectors returned by \texttt{Nii-body} and \texttt{REBOUND} at epoch $t_k$, respectively.

Table~\ref{tab2} summarizes the execution times of \texttt{Nii-body}'s RKF78 integrator and \texttt{REBOUND}'s IAS15 integrator for all the benchmarks.
It gives the RMSE computed with a 100-year output cadence, corresponding to $N = 100$ in Equation \ref{eq:rmse}.
The negligible RMSEs between the two integrators confirm that the trajectories computed by the two codes are identical, as shown in Figure \ref{fig3}.
Table~\ref{tab2} also compares the computational efficiency of the two codes by averaging the execution times across 5 independent runs, demonstrating that the  RKF78 integrator implemented in \texttt{Nii-body} is computationally efficient.
It is important to note that these benchmarks are intended to validate the accuracy and numerical performance of \texttt{Nii-body}'s RKF78 integrator for orbit-fitting problems, and are not meant to recommend it as a general-purpose choice for all long-term $N$-body applications. A direct end-to-end comparison with other existing $N$-body fitting frameworks using the same synthetic Kepler-9 data set lies beyond the scope of the present work. We leave such a comparison to future studies.

\begin{table}
   \caption{Benchmarks between \texttt{Nii-body}'s RKF78  and \texttt{REBOUND}'s IAS15 integrators on the SEJ system over 10,000 years.}
  \footnotesize
  \setlength{\tabcolsep}{15pt}
  \renewcommand{\arraystretch}{0.95}
  \label{tab2}
  \begin{tabular}{cccc} 
    \hline\hline
       $err_{\rm toler}$ & $t_{\texttt{Nii-body}}$  & $t_{\texttt{REBOUND}}$ & RMSE  \\ 
    &  (s) & (s) & (AU)  \\
    \noalign{\smallskip}\hline\noalign{\smallskip}
    $10^{-14}$ & 4.984 & 5.193 & $2.158\times10^{-14}$ \\ 
    $10^{-12}$ & 2.712 & 2.821 & $4.601\times10^{-12}$ \\ 
    $10^{-10}$ & 1.727 & 1.891 & $2.615\times10^{-10}$ \\ 
    $10^{-8}$  & 0.993 & 1.135 & $4.682\times10^{-8}$ \\ 
    \hline
\end{tabular}
\end{table}
\normalsize

\subsection{Astrometric fitting}
\label{sec:kep9peak}

\begin{figure*}[!htb]
    \centering
    \includegraphics[width=1.0\textwidth]{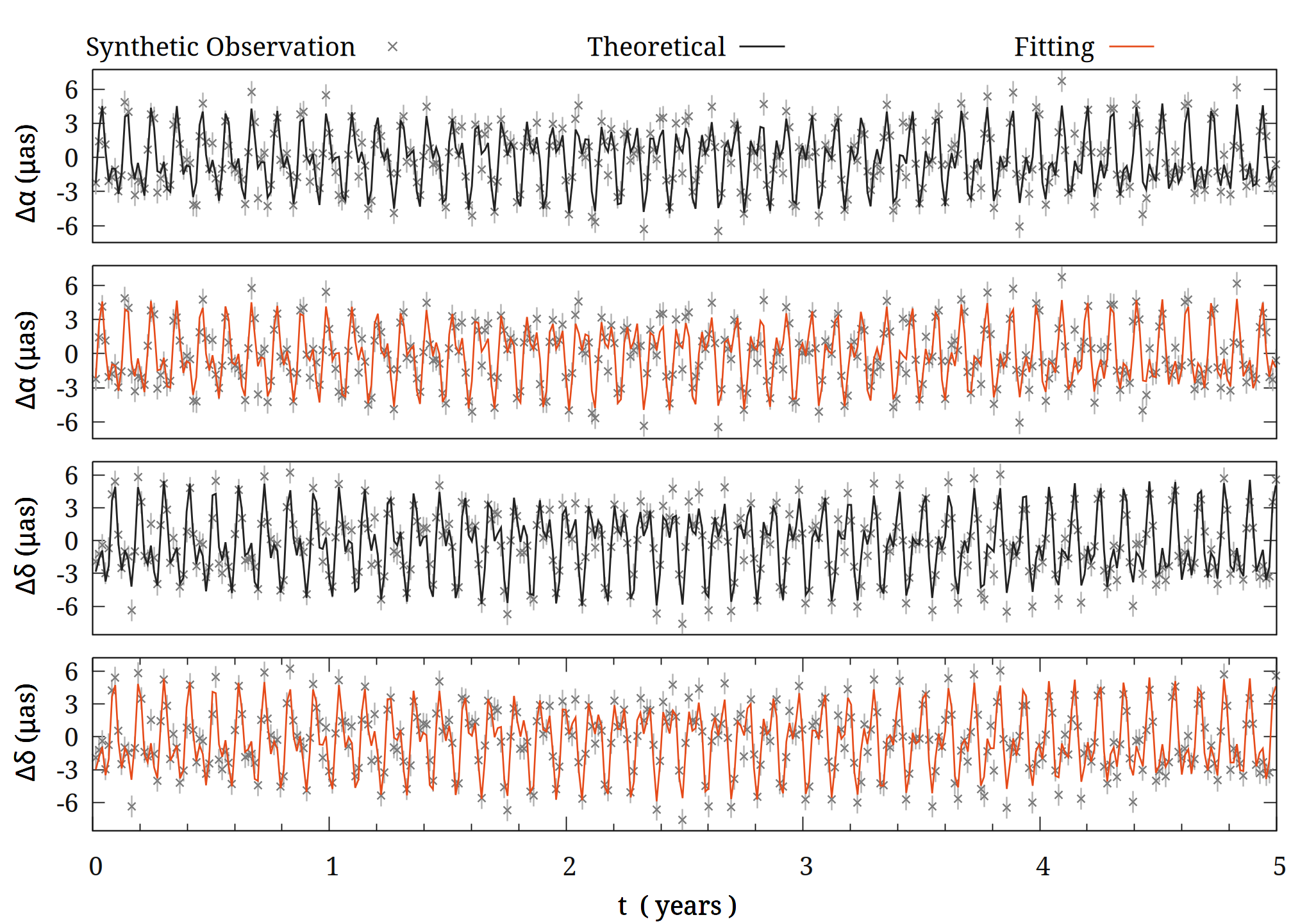}
    \caption{
    The black lines show the theoretical astrometric wobbles of the simulated Kepler-9 star over the five-year period, assuming a distance of 5~pc. 
    The dots and their error bars represent the synthetic observations of the system, generated by adding Gaussian observational noise with a standard deviation of $\sigma = 1~\mu$as on the theoretical astrometric wobbles. 
    The red lines depict the fitted astrometric motion corresponding to the parameters obtained from the N-body MCMC retrieval.
    The close agreement between the black and red lines demonstrates the effectiveness of \texttt{Nii-body}'s N-body fitting approach.
    }

    \label{fig4}
\end{figure*}

Since the primary purpose of \texttt{Nii-body} is orbital fitting of multiplanet systems, we assess its performance using synthetic astrometric signals.
The Kepler-9 system, listed in Table \ref{tab1}, is ideal for testing N-body fitting of planetary gravitational interactions because it hosts two planets in a 2:1 near-resonant configuration.
We therefore selected the first five years of the host star's astrometric time series, as shown by the red curves in Figure~\ref{fig1}, sampled with a uniform cadence of 5 days.
We also added Gaussian observational noise with a standard deviation of 1~$\mu$as to each coordinate at every epoch, corresponding to $\epsilon_x = 1$~$\mu$as in Equation \ref{eqn:lik} , i.e., an identical isotropic uncertainty adopted for both right ascension and declination.
Since this work only aims to be a proof-of-concept validation based on synthetic observation, rather than an analysis of real astrometric measurements, our current model retains only the reflex motion of the host star about the system barycenter, and does not include proper motion, parallax, or the bulk barycentric drift of the system. 
The additional parameters and degeneracies that would arise from these effects are therefore beyond the scope of the present paper, but will be addressed in future applications to real observational data.
Figure~\ref{fig4} shows the theoretical astrometric wobbles and synthetic observations generated assuming a Gaussian observational bias of 1 $\mu\mathrm{as}$ in both right ascension and declination.

Using these synthetic astrometric data, we performed Bayesian orbital fitting by coupling automatic parallel-tempering MCMC sampling with N-body orbital integration. For the Kepler-9 integrations used in the present fitting experiment, we also monitored the relative conservation of the total energy and angular momentum over the five-year fitting window. Here, the total energy and angular momentum are defined as
\begin{equation}
E=\sum_i \frac{1}{2}m_i v_i^2-\sum_{i<j}\frac{m_i m_j}{r_{ij}},
\label{eq:Etot}
\end{equation}
\begin{equation}
\mathbf{L}=\sum_i m_i(\mathbf{r}_i\times\mathbf{v}_i),
\label{eq:Ltot}
\end{equation}
and the corresponding relative drifts are evaluated as
\begin{equation}
\left|\frac{E(t)-E_0}{E_0}\right|,
\qquad
\left|\frac{|\mathbf{L}(t)|-|\mathbf{L}_0|}{|\mathbf{L}_0|}\right|.
\label{eq:drift}
\end{equation}
The maximum values of these two quantities are $1.27\times10^{-13}$ and $1.62\times10^{-14}$, respectively, indicating that no significant secular drift is present at the adopted integration tolerance.

Every model parameter was assigned a uniform prior whose bounds are listed in Table \ref{tab3}.

Due to the robust global-convergence capabilities inherited from the \texttt{Nii-C} MCMC engine~\citep{Jin2024}, the N-body fitting to the synthetic astrometric data can converge from a blind search---no initial guesses required---within 1,000,000 iterations.
A typical MCMC run of this length finishes in $\sim$ 24 hours on an Intel i7-12650H laptop.

\begin{figure*}[!htb]
    \centering
    \includegraphics[width=1.0\textwidth]{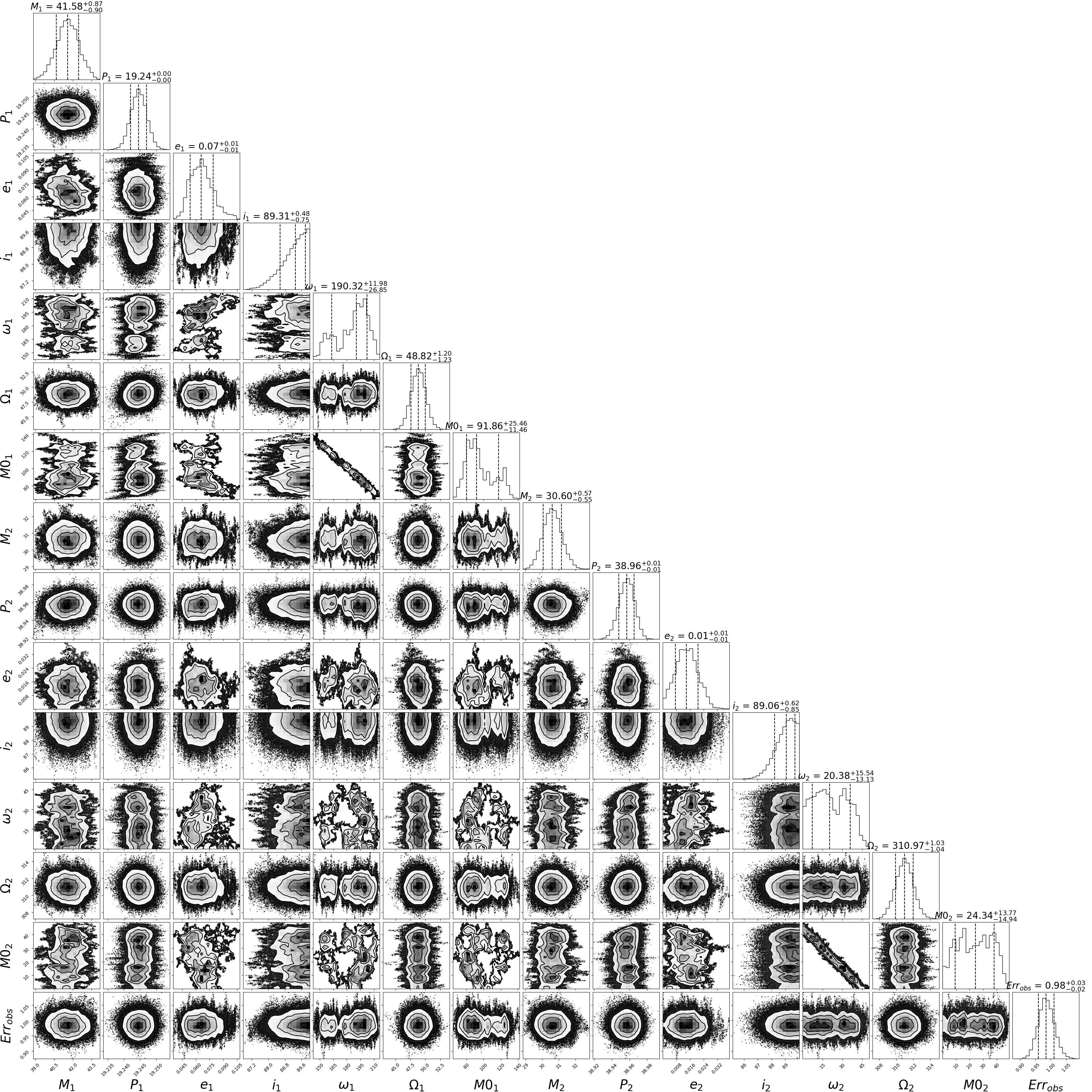}
    \caption{Corner plot displaying the marginal posterior distributions of all 15 parameters obtained from a N-body MCMC fitting run. Note that this corner plot corresponds to Run 8, which converges to a single mode of $\Omega$ and $\omega$. Thus, no post-processing was needed to merge potential bimodal distributions for these parameters. However, this is not always the case for general runs, which can sometimes yield bimodal distributions of $\Omega$ and $\omega$.
    }
    \label{fig5}
\end{figure*}

Figure \ref{fig5} presents the corner plot using the last 500,000 iterations from one N-body fitting.
For a two-planet system, there are a total of 15 system parameters: a planetary mass and six orbital elements for each planet, plus the standard deviation of the Gaussian observational bias.
The posterior means from the MCMC sampling are consistent with the values used to generate the synthetic observational data listed in Table \ref{tab1}.
Note that astrometric data alone cannot uniquely determine the angular parameters $\Omega$ and $\omega$, due to the degeneracy between $\Omega$ and $\Omega+\pi$, or $\omega$ and $\omega+\pi$.
Complementary information---such as radial velocity measurements---can help break this ambiguity.

The fitted astrometric motion corresponding to the parameters obtained from the N-body MCMC retrieval is also presented in Figure \ref{fig4}. 
 The consistency between the theoretical wobbles and the fitted motion demonstrates \texttt{Nii-body}'s N-body fitting capability.

\begin{table}
\caption[Choice of Priors]
{Prior distributions of the planetary orbital parameters.}
\setlength{\tabcolsep}{17pt}
\begin{tabular}{cccc}
\hline
\hline
Parameter & Prior  & Min & Max \\ 
\hline
$P ({\mathrm {days}})$& Uniform & 1.0 & 100 \\ 
&  &  &  \\ 
$M_{\mbox{p}} (M_{\oplus})$ & Uniform & 1.0 &100 \\ 
&  &  &  \\ 
$e$ & Uniform & 0 & 0.5 \\ 
&  &  &  \\ 
$i$ & Uniform & 0 & 90 \\ 
&  &  &  \\ 
$\Omega$ & Uniform & 0 & $2\pi$ \\ 
&  &  &  \\ 
$\omega$ & Uniform & 0 & $2\pi$ \\ 
&  &  &  \\ 
$M_{\mathrm 0}$ & Uniform & 0 & $2\pi$ \\ 
&  &  &  \\ 
$\epsilon_{x} (\mu\mbox{as})$ & Uniform & 0.1 & $10$\\

\hline
\end{tabular}
\label{tab3}
\end{table}
\normalsize

To test the robustness of the N-body fitting, we conducted eight automatic parallel tempering MCMC runs using \texttt{Nii-body}.
All control parameters for the eight runs were set to the same values, with the only difference being the initial random seeds for the MCMC. 
Table~\ref{tab4} summarizes the major system parameters of the Kepler-9 system retrieved from eight  N-body fitting MCMC runs.
The results show that 6 of the 8 runs --- a 75\% recovery rate --- successfully retrieved the planetary parameters of the Kepler-9 system.
Run 4 did not achieve convergence; it exhibits a large lower uncertainty in the fitted orbital period of the first planet, reported as $19.23^{+0.01}_{-12.48}$.
This large lower error bar arises because there is another high-density island at longer periods in the posterior distribution, and the MCMC in Run 4 has not yet reached global convergence. A plausible explanation is that the uniform time sampling used in our model to generate synthetic astrometric data can introduce harmonics, leading to local maxima in the likelihood function.
Another failed run is Run 3. Although it correctly recovers the planetary mass and orbital period, it yields a higher eccentricity and lower inclination, indicating a potential degeneracy between these two parameters. The higher-eccentricity, lower-inclination combination found in Run 3 yields a marginally lower posterior probability, and forward integration shows that this configuration is long-term unstable.
Runs 3 and 4 show that, while \texttt{Nii-body} reliably retrieves planetary masses and orbital periods, independent MCMC chains remain essential to confirm full global convergence.

\begin{table*}[!t]
\caption[]{Principal parameters of the synthetic Kepler-9 system retrieved from eight $N$-body fittings (upper block) and eight Keplerian-superposition fittings (lower block). Six of the eight $N$-body runs successfully recovered the planetary parameters, whereas Run 4 did not achieve convergence and Run 3 converged to a degenerate but dynamically unstable solution. None of the Keplerian-superposition fittings recovered the true parameters.}
  \label{tab4}
  \centering
  \scriptsize
  \setlength{\tabcolsep}{7.5pt}
  \renewcommand{\arraystretch}{1.3}

  \begin{tabular}{c|ccccccccc}
  \hline
  \hline
  Case & $M_1$ ($M_\oplus$) & $P_1$ (days) & $e_1$ & $i_1$ &
       $M_2$ ($M_\oplus$) & $P_2$ (days) & $e_2$ & $i_2$ & $Err_{\rm obs}$ \\
  \hline    
  Input truth & $41.71$ & $19.25$ & $0.06$ & $88.94$ & $30.79$ & $38.94$ & $0.07$ & $89.18$ & $1.00$ \\
  \hline
  Run 1 & $41.76^{+0.87}_{-0.87}$ & $19.24^{+0.00}_{-0.00}$ & $0.07^{+0.01}_{-0.01}$ & $89.29^{+0.51}_{-0.81}$ & $30.61^{+0.55}_{-0.54}$ & $38.96^{+0.01}_{-0.01}$ & $0.02^{+0.00}_{-0.00}$ & $89.02^{+0.65}_{-0.87}$ & $0.98^{+0.03}_{-0.03}$ \\
  Run 2 & $41.85^{+0.84}_{-0.87}$ & $19.24^{+0.00}_{-0.00}$ & $0.06^{+0.01}_{-0.01}$ & $89.32^{+0.48}_{-0.79}$ & $30.58^{+0.55}_{-0.55}$ & $38.96^{+0.01}_{-0.01}$ & $0.02^{+0.01}_{-0.01}$ & $89.02^{+0.64}_{-0.89}$ & $0.99^{+0.03}_{-0.03}$ \\
  Run 3 & $41.71^{+1.23}_{-1.10}$ & $19.22^{+0.00}_{-0.00}$ & $0.40^{+0.00}_{-0.00}$ & $85.78^{+0.36}_{-0.58}$ & $29.88^{+0.62}_{-0.60}$ & $39.05^{+0.01}_{-0.01}$ & $0.00^{+0.00}_{-0.00}$ & $86.19^{+1.18}_{-0.53}$ & $1.13^{+0.03}_{-0.03}$ \\
  Run 4 & $41.77^{+42.05}_{-1.49}$ & $19.24^{+0.01}_{-12.48}$ & $0.08^{+0.01}_{-0.01}$ & $89.30^{+0.51}_{-0.86}$ & $30.58^{+0.57}_{-0.53}$ & $38.97^{+0.01}_{-0.03}$ & $0.02^{+0.01}_{-0.01}$ & $89.05^{+0.64}_{-0.86}$ & $0.98^{+0.03}_{-0.02}$ \\
  Run 5 & $41.77^{+0.85}_{-0.86}$ & $19.24^{+0.00}_{-0.00}$ & $0.07^{+0.01}_{-0.01}$ & $89.31^{+0.48}_{-0.76}$ & $30.60^{+0.55}_{-0.53}$ & $38.98^{+0.01}_{-0.02}$ & $0.02^{+0.03}_{-0.02}$ & $88.99^{+0.67}_{-0.88}$ & $0.98^{+0.03}_{-0.03}$ \\
  Run 6 & $41.77^{+0.88}_{-0.88}$ & $19.24^{+0.00}_{-0.00}$ & $0.07^{+0.01}_{-0.01}$ & $89.29^{+0.51}_{-0.81}$ & $30.57^{+0.55}_{-0.54}$ & $38.97^{+0.01}_{-0.01}$ & $0.01^{+0.01}_{-0.01}$ & $89.02^{+0.65}_{-0.88}$ & $0.98^{+0.03}_{-0.03}$ \\
  Run 7 & $41.92^{+0.82}_{-0.82}$ & $19.24^{+0.00}_{-0.00}$ & $0.06^{+0.02}_{-0.01}$ & $89.37^{+0.45}_{-0.77}$ & $30.51^{+0.54}_{-0.55}$ & $38.97^{+0.01}_{-0.01}$ & $0.03^{+0.02}_{-0.02}$ & $89.19^{+0.53}_{-0.82}$ & $1.00^{+0.03}_{-0.03}$ \\
  Run 8 & $41.58^{+0.87}_{-0.90}$ & $19.24^{+0.00}_{-0.00}$ & $0.07^{+0.01}_{-0.01}$ & $89.31^{+0.48}_{-0.75}$ & $30.60^{+0.57}_{-0.55}$ & $38.96^{+0.01}_{-0.01}$ & $0.01^{+0.01}_{-0.01}$ & $89.06^{+0.62}_{-0.85}$ & $0.98^{+0.03}_{-0.02}$ \\
  \hline\noalign{\smallskip}
  Run 1 & $48.47^{+1.91}_{-30.43}$ & $6.75^{+32.16}_{-0.00}$ & $0.06^{+0.02}_{-0.04}$ & $67.53^{+12.50}_{-25.14}$ & $18.50^{+6.32}_{-0.56}$ & $38.91^{+0.01}_{-19.66}$ & $0.04^{+0.03}_{-0.02}$ & $29.87^{+6.28}_{-28.27}$ & $0.98^{+0.03}_{-0.03}$ \\
  Run 2 & $47.52^{+2.53}_{-29.51}$ & $6.76^{+32.16}_{-0.00}$ & $0.05^{+0.02}_{-0.03}$ & $42.45^{+6.24}_{-0.05}$ & $18.78^{+6.11}_{-0.81}$ & $38.90^{+0.01}_{-19.65}$ & $0.04^{+0.04}_{-0.02}$ & $58.09^{+0.04}_{-28.25}$ & $0.98^{+0.03}_{-0.03}$ \\
  Run 3 & $48.25^{+1.92}_{-23.81}$ & $6.75^{+12.50}_{-0.00}$ & $0.06^{+0.01}_{-0.01}$ & $17.30^{+15.68}_{-0.04}$ & $18.58^{+59.49}_{-0.61}$ & $38.91^{+0.01}_{-34.47}$ & $0.02^{+0.02}_{-0.01}$ & $14.17^{+31.41}_{-0.03}$ & $0.98^{+0.03}_{-0.03}$ \\
  Run 4 & $24.59^{+44.24}_{-6.33}$ & $19.25^{+19.66}_{-14.82}$ & $0.06^{+0.02}_{-0.04}$ & $17.24^{+47.15}_{-6.25}$ & $18.32^{+31.05}_{-0.43}$ & $38.91^{+0.01}_{-32.15}$ & $0.04^{+0.05}_{-0.02}$ & $51.83^{+0.04}_{-25.14}$ & $0.98^{+0.03}_{-0.02}$ \\
  Run 5 & $24.44^{+0.60}_{-0.18}$ & $19.25^{+19.66}_{-0.00}$ & $0.06^{+0.02}_{-0.03}$ & $45.52^{+37.72}_{-18.83}$ & $18.28^{+30.88}_{-0.39}$ & $38.91^{+0.01}_{-32.16}$ & $0.03^{+0.03}_{-0.02}$ & $80.07^{+9.47}_{-65.91}$ & $0.98^{+0.03}_{-0.03}$ \\
  Run 6 & $49.08^{+21.39}_{-24.55}$ & $6.75^{+12.50}_{-2.79}$ & $0.07^{+0.01}_{-0.01}$ & $79.96^{+6.29}_{-50.27}$ & $18.13^{+0.32}_{-0.32}$ & $38.91^{+0.01}_{-0.01}$ & $0.02^{+0.02}_{-0.01}$ & $32.99^{+34.55}_{-6.29}$ & $0.98^{+0.03}_{-0.03}$ \\
  Run 7 & $66.13^{+5.07}_{-47.69}$ & $3.97^{+34.93}_{-0.00}$ & $0.06^{+0.02}_{-0.03}$ & $36.12^{+6.33}_{-15.70}$ & $18.23^{+5.89}_{-0.37}$ & $38.91^{+0.01}_{-19.66}$ & $0.02^{+0.03}_{-0.01}$ & $39.28^{+12.59}_{-28.26}$ & $0.98^{+0.03}_{-0.03}$ \\
  Run 8 & $24.14^{+0.81}_{-6.06}$ & $19.25^{+19.66}_{-0.00}$ & $0.06^{+0.02}_{-0.04}$ & $20.42^{+21.99}_{-15.71}$ & $18.44^{+53.07}_{-0.52}$ & $38.91^{+0.01}_{-34.94}$ & $0.04^{+0.03}_{-0.03}$ & $32.97^{+25.15}_{-12.55}$ & $0.98^{+0.03}_{-0.03}$ \\
  \hline
  \end{tabular}
  \end{table*}

To verify the necessity of using N-body MCMC fitting, we repeated the retrieval exercise with eight independent automated parallel tempering MCMC runs employing a Keplerian superposition model~\citep{Jin2024}.
The lower half of Table \ref{tab4} summarizes the resulting posteriors: none of the eight MCMC runs recovered the true parameters, and the extremely broad credible intervals indicate a failure to converge.
Therefore, for systems like Kepler-9, where planets b and c reside in a 2:1 mean motion resonance, the Keplerian superposition model is inadequate, and N-body fitting is essential for accurately determining the system parameters.
The explanation is straightforward: the Keplerian superposition model ignores planet–planet perturbations that are dynamically dominant in compact, interacting architectures, whereas the N-body forward model incorporates them self-consistently.

\section{Summary}
\label{sec:sum}

We introduce  \texttt{Nii-body}, an open source N-body MCMC fitting code  primarily developed for the Bayesian retrieval of orbital parameters in  multiplanet systems.
The code implements an RKF78 integrator to track the gravitational interactions within a planetary system with high accuracy and practical computational performance.
Equipped with the automatic parallel tempering MCMC engine inherited from \texttt{Nii-C}~\citep{Jin2024}, \texttt{Nii-body} provides an effective framework for Bayesian retrieval of orbital parameters in dynamically interacting multiplanet systems.

We tested the  N-body orbital fitting performance of \texttt{Nii-body} using  synthetic astrometric observations modeled after the Kepler-9 two-planet system, assuming a distance of 5 pc.
\texttt{Nii-body} reached convergence from random starting points in the full 15-dimensional parameter space in 75\% of the runs within approximately 24 hours on an i7-12650H laptop. This performance indicates that N-body orbital retrieval is computationally practical for the synthetic test cases considered in this work.
Our test case with the Kepler-9 system further demonstrates that, for compact multiplanet systems exhibiting strong dynamical interactions, particularly those near mean motion resonances, N-body orbital integration is mandatory to accurately capture the mutual gravitational perturbation and, consequently, to precisely determine the orbital elements.

In this paper, we focus solely on astrometric observations of multiplanet systems. However, the fitting framework can be readily extended to enable joint dynamical modeling of astrometry, RV, and TTV data, allowing for more comprehensive studies of the dynamical evolution of real exoplanetary systems.
A further important direction involves implementing explicit dynamical stability constraints within MCMC sampling to guarantee long-term stability of fitted orbital configurations.
We reserve these developments for future work and intend to incorporate them in subsequent releases.

Moreover, \texttt{Nii-body} serves as a general retrieval tool for 
N-body gravitational interactions across various disciplines in astrophysics. With additional pipelines for synthetic observations, \texttt{Nii-body} can facilitate the fitting of a broad range of N-body gravitational systems, spanning stellar kinematics to galactic-scale dynamics.

\section*{acknowledgements}

This work is supported by National Natural Science Foundation of China (Grant No. 11973094, 12033010, 12573076), the incubation programme for recruited talents (2023GFXK153), and the doctoral start-up funds from Anhui Normal University. 

\section*{Author Contributions}
Sheng Jin, Dong-Hong Wu, and Shang-Fei Liu conceived the ideas. Dong-Hong Wu designed the N-body integration scheme. Hongfei Jia and Sheng Jin wrote the code, performed the numerical experiment, and wrote the manuscript. Shang-Fei Liu validated the N-body benchmarks.
All authors read and approved the final manuscript.
 
\section*{Declaration of Interests}
The authors declare no competing interests.

\normalsize
\clearpage

\bibliographystyle{ati} 
\bibliography{ati}      

\label{lastpage}

\end{document}